\documentclass[12pt,titlepage]{article}

\usepackage{rotating,epsfig}

\usepackage{graphicx}

\usepackage{amsmath}
\usepackage{amssymb,bm}

\usepackage{ulem}

\usepackage{psfrag}
\usepackage{cancel}
\usepackage{graphicx}

\usepackage{color}

\newcommand{\yy}{\mbox{$\mathbf y$}}

\newcommand{\be}{\mbox{\boldmath $\beta$}}

\newcommand{\vart}{\mbox{\boldmath $\vartheta$}}

\newcommand{\bb}{\mbox{$\mathbf b$}}
\newcommand{\ep}{\mbox{\boldmath $\epsilon$}}

\newcommand{\vecbeta}{{\bm{\beta}}}

\newcommand{\vecmu}{\bm{\mu}}

\newcommand{\vecrho}{{\bm{\rho}}}

\newcommand{\vecpsi}{\bm{\psi}}

\newcommand{\matA}{\mathbf{A}}
\newcommand{\matB}{\mathbf{B}}
\newcommand{\matD}{\mathbf{D}}
\newcommand{\matE}{\mathbf{E}}

\newcommand{\matI}{\mathbf{I}}

\newcommand{\matO}{\mathbf{O}}

\newcommand{\matS}{\mathbf{S}}
\newcommand{\matV}{\mathbf{V}}
\newcommand{\matU}{\mathbf{U}}
\newcommand{\matW}{\mathbf{W}}
\newcommand{\matX}{\mathbf{X}}
\newcommand{\matZ}{\mathbf{Z}}

\newcommand{\matPsi}{\mathbf{\Psi}}

\newcommand{\E}{\mathrm E}
\newcommand{\tr}{\mathrm{tr}}
\newcommand{\vct}{\mathrm{vec}}
\newcommand{\vcth}{\mathrm{vech}}
\newcommand{\partiald}[2]{{\displaystyle\frac{\partial #1}{\partial #2}}}
\newcommand{\half}[1]{\frac{#1}{2}}

\makeatletter
\def\bbordermatrix#1{\begingroup \m@th
  \@tempdima 4.75\p@
  \setbox\z@\vbox{%
    \def\cr{\crcr\noalign{\kern2\p@\global\let\cr\endline}}%
    \ialign{$##$\hfil\kern2\p@\kern\@tempdima&\thinspace\hfil$##$\hfil
      &&\quad\hfil$##$\hfil\crcr
      \omit\strut\hfil\crcr\noalign{\kern-\baselineskip}%
      #1\crcr\omit\strut\cr}}%
  \setbox\tw@\vbox{\unvcopy\z@\global\setbox\@ne\lastbox}%
  \setbox\tw@\hbox{\unhbox\@ne\unskip\global\setbox\@ne\lastbox}%
  \setbox\tw@\hbox{$\kern\wd\@ne\kern-\@tempdima\left[\kern-\wd\@ne
    \global\setbox\@ne\vbox{\box\@ne\kern2\p@}%
    \vcenter{\kern-\ht\@ne\unvbox\z@\kern-\baselineskip}\,\right]$}%
  \null\;\vbox{\kern\ht\@ne\box\tw@}\endgroup}
\makeatother

\begin{document}

\begin{center}

\centerline{\bf The Effective Sample Size in Bayesian Information Criterion}

\centerline{\bf for Level-Specific Fixed and Random Effects Selection in a Two-Level Nested Model}

\vspace{1.5in}

\centerline{Sun-Joo Cho}

\centerline{Vanderbilt University}

\vspace{.3in}

\centerline{Hao Wu}

\centerline{Vanderbilt University}

\vspace{.3in}

\centerline{Matthew Naveiras }

\centerline{Vanderbilt University}

\end{center}


\vspace{.3in}

\centerline{March 6, 2022}

\vspace{1.0in}

\noindent{\it Note.} The first and second authors contributed equally to this work.

\vspace{1.0in}

\noindent{Correspondence should be sent to\\

\noindent E-Mail: sj.cho@vanderbilt.edu      \break
\noindent Phone: 615-322-8409 \break
\noindent Website: $http://www.vanderbilt.edu/psychological_sciences/bio/sun-joo-cho$   }

\clearpage
\newpage

\begin{center}

\centerline{\bf The Effective Sample Size in Bayesian Information Criterion} 

\centerline{\bf for Level-Specific Fixed and Random Effects Selection in a Two-Level Nested Model}

\end{center}

%

\baselineskip 24pt

\noindent Popular statistical software provides Bayesian information criterion (BIC) for multilevel models or linear mixed models.
However, it has been observed that the combination of statistical literature and software
documentation has led to discrepancies in the formulas of the BIC and uncertainties
of the proper use of the BIC in selecting a multilevel model with respect to level-specific fixed and random effects.
These discrepancies and uncertainties result from different specifications of sample size in the BIC's penalty term for multilevel models.
In this study, we derive the BIC's penalty term for level-specific fixed and random effect selection in a two-level nested design.
In this new version of BIC, called BIC$_E$, this penalty term  is decomposed into two parts if the random effect variance-covariance matrix has full rank: (a) a term with the log of average sample size per cluster whose multiplier involves the overlapping number of
dimensions between the column spaces of the random and fixed effect design matrices and (b) the total number of parameters times the log of the total number of clusters.
Furthermore, we study the behavior of BIC$_{E}$ in the presence of redundant random effects.
The use of BIC$_{E}$ is illustrated with a textbook example data set and a numerical demonstration shows that the derived formulae adheres to empirical values.


\clearpage
\newpage

\section{Introduction}

Multilevel models (MLMs; e.g., Goldstein, 2003) or linear mixed models (Laird \& Ware, 1982) are a general class of modeling framework to describe
the relationship between the response and covariates for clustered data including repeated measures and nested designs.
Selecting fixed and random effects is an important step in the applications of the MLM.
When competing MLMs are compared in model selection based on the maximum likelihood (ML) method, the literature suggests using the likelihood ratio test (LRT) or the deviance test to compare the fit of the two competing MLMs which differ in the fixed and/or random effects (e.g., Goldstein, 2003, p. 24, pp. 35--36; Snijders \& Bosker, 1999, pp. 88--90).
According to the survey of Whittaker and Furlow (2009) regarding the use of model selection methods for the MLMs, the LRT is a dominant model selection method in the applications of the MLMs.

Using the LRT, one can make a statistical decision by comparing the likelihood under two competing models against the critical value, which can easily be found based on the known null distribution.
However, difficulties of using the LRT for MLMs have been noted in the literature.
First, the LRT is mainly for the comparison of two nested models, although a line of research on the LRT for non-nested models (e.g., Merkle, You, \& Preacher, 2016; Vuong, 1989) also exists.
Second, the LRT statistic does not follow a simple chi-square distribution when the random effect variance-covariance matrix has a boundary value on the model parameter space (Self \& Liang, 1987; Stram \& Lee, 1994\footnote{Stram and Lee (1994) incorrectly specified constraints in their derivation, which they later noted (Stram \& Lee, 1995).}, 1995; Molenbergh \& Verbeke, 2007).
Third, a LRT does not quantity the degree to which a model is better than another model.
Based on the LRT results, we can only conclude that the two comparison models fit equally well or that the more complex model fits better.
Fourth, small changes that are too small to be of practical importance always becomes significant with large enough sample sizes (e.g., Jones [2011]).
Because most null hypotheses are rejected in large sample sizes, hypothesis tests often suggest complex models (Weaklim, 1999).

In addition to the LRT, information criteria such as the Bayesian information criterion (BIC; Schwarz, 1978) have been used for the MLMs (e.g., Hamaker, van Hattum, Kuiper, \& Hoijtink, 2011; McCoach \& Black, 2008; Whittaker \& Furlow, 2009).
Most software for the MLMs, including SPSS (IBM Corp., 2020), the MIXED procedure in SAS (SAS Institute Inc., 2015), R {\tt lme4} library (Bates, M{\"a}chler, Bolker, \& Walker, 2015), and Mplus (Muth\'{e}n \& Muth\'{e}n, 1998--2015), provide BIC and/or its sample-size modification as a part of its output.
The BIC allows for the comparisons of two or more competing models, whether or not they are nested, and it quantifies the degree to which a given model represents an improvement over the other competing models (Burnham \& Anderson, 2002).
The BIC is relatively easy to calculate and under certain conditions the difference between two BICs is a rough approximation to the logarithm of the Bayes factor which requires evaluation of prior distributions (Berger, Ghosh, \& Mukhopadhyay, 2003; Kass \& Raftery, 1995).
BIC does not constitute a statistical test of the difference in the competing models and model selection is made based on ranking of the BIC values among competing models.

However, it is challenging to use the BIC for the MLM because the sample size in the BIC's penalty term is not clear.
In the literature, it was noted that the sample size in the BIC's penalty term is not well-defined for the dependent observations in clustered data.
For the MLM, the sample size in BIC calculation can be a total sample size, the number of clusters, cluster sizes, or the weighted average of the total sample size and cluster sizes (e.g., Hamaker, van Hattum, Kuiper, \& Hoijtink, 2011, p. 249; McCoach \& Black, 2008, p. 253).
We found inconsistent use of sample size in the BIC's penalty term across software.
SPSS, R, and Mplus use the total sample size for all kinds of MLMs, whereas the MIXED procedure in SAS (SAS Institute Inc., 2015, p. 6064) use the number of levels of the first random effect specified in a RANDOM statement for the sample size (e.g., the number of clusters for a two-level random intercept model).
Furthermore, MLM textbooks also differ in their recommendations.
For example, Hox, Moerbeek, and van de Schoot (2018, p. 39) and Goldstein (2003, p. 37) noted that the total number of higher level units in MLM is often used as an approximation of the effective sample size. However, Snijders and Bosker (2012, p. 202) presented BIC with the total sample size in MLM applications.
When the sample size in the BIC's penalty term is not correctly defined for the MLMs, different calculation of BIC results in different model selection results.

For the MLM, observations within a cluster tend to be dependent, which impacts the model selection with the BIC by reducing the sample size.
Several researchers attempted to calculate an {\it effective sample size} for dependent observations.
Pauler (1998) presented the effective sample size for choosing fixed effects in normal linear mixed models (or random intercept models) and her main idea for the modified BIC was to have a different penalty term per parameter.
Raftery (1995) defined the effective sample size for a linear regression model and analysis of variance, a logistic regression model, a log-linear model, an event-history model, and a structural equation model.
Berger, Ghosh, and Mukhopadhyay (2003, p. 243) and Kass and Raftery (1995, p. 779) noted that the effective sample size can be derived as the scalar for an approximation of the information matrix.
However, these previous studies cannot be applied directly for MLM.
Jones (2011) derived the effective sample size in the BIC's penalty term for a linear mixed model as the sum of the elements of the inverse of each cluster's correlation matrix (of the response) across clusters (Equation 3 in Jones [2011]). Despite the author showing that the effective sample size is a function of the intraclass correlation (ICC) (Equation 6 in Jones [2011]), he did not show the effective sample size in the context of selecting the level-specific fixed and random effects (random intercept and random slope) of the linear mixed model.
Delattre, Lavielle, and Poursat (2014) derived BIC for mixed-effects models in which the effective sample size is the number of clusters for random effects but is the total sample size for fixed effects. However, they did not provide the effective sample size for level-specific fixed effects.
Recently, Lorah and Womack (2019) showed via a simulation study that BIC produces appropriate model selection behavior regarding {\it fixed} effects in the case that the effective sample size is the number of clusters for between-cluster fixed effects and but is the total sample size for within-cluster fixed effects in a two-level random intercept model. Although findings in Delattre et al. (2014) and Lorah and Womack (2019) suggest the importance of appropriate the effective sample size for either level-specific fixed effects or random effects, these studies did not show the derivation of the effective sample size in the selection of level-specific fixed and random effects (random intercept and random slope).

Thus, the purpose of this paper is to derive the BIC's penalty term for the level-specific fixed and random effect selection in MLM.
The focus is not to argue in favor of BIC against LRT or other information criteria.
Specifically, we analyze the large sample behavior of the information matrix in the MLM for two-level nested data (e.g., students nested within schools).

The remainder of this paper is organized as follows.
In Section 2, the BIC and Laplace approximation as its theoretical basis are reviewed.
In Section 3, we derive the BIC for a two-level nested model when the variance-covariance matrix of random effects has full rank.
In Section 4, we derive the BIC for models with redundant random effects.
In Section 5, the behavior of BIC is demonstrated in various multilevel designs.
In Section 6, the use of BIC is illustrated using a textbook example.
In Section 7, we end with a summary and a discussion.

\section{Bayesian Information Criterion (BIC)}

Schwarz (1978) derived the BIC as an asymptotic approximation of the Bayesian marginal probability of a candidate model $M$:
\begin{equation}
\log f(\yy|M) = \log f(\yy|\widehat{\vart},M) - \frac{K}{2}\log N + O_p(1),  \label{Schwarz}
\end{equation}
where $\yy$ is data, $\vart$ is the vector of parameters for model $M$, $f(\yy|\widehat{\vart},M)$ is the likelihood of data $\yy$ evaluated at the ML estimate (MLE) $\widehat{\vart}$, $K$ is the number of parameters for model $M$,
and $N$ is the sample size. For a large sample size, the order term $O_p(1)$ can be dropped in Equation~\ref{Schwarz}.
The BIC can also be presented (for a large sample size) as:
\begin{equation}
{\rm BIC} = -2\log f(\yy|\widehat{\vart},M) + K\log N.  \label{bicc}
\end{equation}

The original derivation in Schwarz (1978) was ``for the case of independent, identically distributed (i.i.d.) observations, and linear models'', under the assumption that the likelihood belongs to an exponential family.
It has been noted that BIC can be used for observations not necessarily identically distributed (e.g., Pauler, 1998; Stone, 1979).
Below, the notation for the candidate model $M$ as it appears in Equation~\ref{bicc} is omitted for reasons of simplicity.

The BIC can be derived based on Laplace approximation (e.g., Raftery, 1995; Kass \& Vaidyanathan, 1992), which
is also valid for mixed models (e.g., Wolfinger, 1993). In the Laplace approximation method, the log likelihood,  $g(\vart)=\log f(\yy|\vart)$, is expanded as a quadratic Taylor series about the MLE $\hat{\vart}$ under regularity conditions (de Bruijn, 1970, sec. 4.4; Tierney \& Kadane, 1986):
\begin{equation}
	g(\vart) = g(\hat{\vart}) - \frac{1}{2}(\vart - \hat{\vart})^{T}\matA(\vart - \tilde{\vart}) + o_p(1),\label{eq:Taylorg}
\end{equation}
where $\matA=-\E g''(\hat{\vart})$ is the information matrix, and $g''(\hat{\vart})$ is the Hessian matrix.
We note that $g'(\hat{\vart})=0$, so the linear term is not present.
the marginal probability $f(\yy)$ can be approximated as (Kass \& Vaidyanathan, 1992)
\begin{align*}
	f(\yy) = \int \exp[g(\vart)] \pi (\vart) d\vart
	= f(\yy|\hat\vart)\pi(\hat\vart)(2\pi)^{K/2}|\matA|^{-1/2}\left\{1+O_p(1/n)\right\},\label{eq3}
\end{align*}
where $\pi(\vart)$ is the prior density.

Taking the logarithm leads to
\begin{equation}
	\log f(\yy) = \log f(\yy|\hat\vart) + \log\pi(\hat\vart) + \frac{K}{2}\log(2\pi) - \frac{1}{2}\log|\matA|+o_p(1). \label{eqq4}
\end{equation}
When observations are i.i.d., $\log|\matA| = K\log(N)+O_p(1)$, leading to BIC as in Equation~\ref{Schwarz}.

\section{BIC for a Two-Level Nested Model}

For two-level nested data, the response vector $\yy_{j}$ is composed of the response $y_{ji}$
from observation $i$ ($i=1,\ldots,n_{j}$) in cluster $j$ ($j=1,\ldots, J$).
The standard form of the variance component model (Laird \& Ware, 1982) is
\begin{equation}
\yy_{j}=\matX_{j}\be + \matZ_{j}\bb_{j} + \ep_{j}, \label{model}
\end{equation}
where
$\matX_{j}$ is ($n_{j} \times p$) design matrix for fixed effects,
$\matZ_{j}$ is ($n_{j} \times q$) design matrix for random effects,
$\be$ is a $p\times 1$ vector of fixed effects,
$\bb_{j}$ is a $q\times 1$ vector of random effects distributed as $\bb_{j} \sim MVN({\bf 0},\matPsi_{j})$ (where $\matPsi_{j}$ is a variance-covariance matrix of random effects), and
$\ep_{j}$ is a $n_{j}\times 1$ vector of residuals distributed as $\ep_{j} \sim N(0, \sigma_{j}^{2}I_{n_{j}})$ where $I_{n_{j}}$ is an identity matrix of size $n_{j}$. In most applications of the MLMs, it is often assumed that all $\matPsi_{j}$ and $\sigma_{j}^{2}$ stay the same across all clusters (e.g., de Leeuw \& Meijer, 2007). Thus, in this study, we assume $\bb_{j} \sim MVN({\bf 0},\matPsi)$ and $\ep_{j} \sim N(0, \sigma^{2}I_{n_{j}})$ (homoscedasticity).
The total sample size is denoted by $N$, calculated as $N=nJ$ for a balanced design ($n_j=n$) and $N=\sum_{j=1}^{J}n_{j}$ for an unbalanced design.
The average cluster size is denoted by $\bar{n}=N/J$.
We also define $\vart = [\be',\vcth'(\matPsi),\sigma^{2}]'$ as the vector of parameters.

For $J$ independent clusters, the matrix $\matA$ in Equation~\ref{eq:Taylorg} can be expressed as
\begin{equation}
\matA=\sum\matI_{j}(\hat{\vart}),
\end{equation}
where $\matI_{j}(\hat{\vart})$ is the Fisher information for cluster $j$.

The mean vector $\vecmu_j$ and the covariance matrix $\matV_j$ for the distribution of $\yy_j$ are given by
\begin{equation}\label{eq: MeanCov}
	\vecmu_j= \matX_j\vecbeta\quad\mbox{and}\quad
	\matV_j= \matZ_j\matPsi\matZ_j'+\sigma^2 I_{n_j}
\end{equation}

When no parameter appears in both the mean and covariance structures, the information matrix of a multivariate normal model is block diagonal, with a block $\matI_{\vecbeta\vecbeta}$ for the mean structure parameters (here the fixed effects $\vecbeta$) and a second block $\matI_{\rho\rho}$ for the covariance structure parameters (here the residual variance $\sigma^{2}$ and the non-duplicated elements in random effect variance-covariance matrix $\vcth'(\matPsi)$; we define $\vecrho = [\vcth'(\matPsi), \sigma^2]'$). The log-determinant of the information matrix can be expressed as
\begin{equation}\label{eq: logdetA}
	\log|\matA| = \log|\matI_{\vecbeta\vecbeta}| + \log|\matI_{\rho\rho}|
\end{equation}
Below we investigate the two blocks separately.


\subsection{The Information Matrix of Fixed Effects}

We define:
\begin{equation}
	\matS_{XX,n_j} = \matX_j'\matX_j/n_j\quad\quad
	\matS_{XZ,n_j} = \matX_j'\matZ_j/n_j\quad\quad
	\matS_{ZZ,n_j} = \matZ_j'\matZ_j/n_j
\end{equation}
Note that while the number of rows of $\matX_j$ and $\matZ_j$ grows with $n_j$, the sizes of the matrices defined above remain constant and their elements are assumed to remain bounded.

The block of $\matI_j$ corresponding to the fixed effects $\vecbeta$ is given by
\begin{equation}\label{eq: Ifixed}
	\matI_{\vecbeta\vecbeta,j}
	= \matX_j'\matV^{-1}_j\matX_j
\end{equation}

Here we note that
\begin{align}
	\matV_j^{-1} &= \frac{1}{\sigma^2}\left\{I_{n_j} - \matZ_j(\sigma^2\matPsi^{-1}+\matZ_j'\matZ_j)^{-1}\matZ_j'\right\}
	= \frac{1}{\sigma^2}\left\{I_{n_j} - \frac{1}{n_j}\matZ_j\left[\frac{\sigma^2}{n_j}\matPsi^{-1}+\matS_{ZZ,n_j}\right]^{-1}\matZ_j'\right\}\\
	&= \frac{1}{\sigma^2}\left\{I_{n_j} - \frac{1}{n_j}\matZ_j\matS_{ZZ,n_j}^{-1}\matZ_j'\right\}
	+\frac{1}{n_j^2}\matZ_j\matS_{ZZ,n_j}^{-1}\left[\frac{\sigma^{2}}{n_j}\matS_{ZZ,n_j}^{-1}+\matPsi\right]^{-1}\matS_{ZZ,n_j}^{-1}\matZ_j'
\end{align}

and Equation~\ref{eq: Ifixed} becomes
\begin{equation*}
	\matI_{\vecbeta\vecbeta,j}= \frac{n_j}{\sigma^2}\left\{\matS_{XX,n_j} - \matS_{XZ,n_j}\matS_{ZZ,n_j}^{-1}\matS_{ZX,n_j}\right\}
	+\matS_{XZ,n_j}\matS_{ZZ,n_j}^{-1}\left[\frac{\sigma^{2}}{n_j}\matS_{ZZ,n_j}^{-1}+\matPsi\right]^{-1}\matS_{ZZ,n_j}^{-1}\matS_{ZX,n_j}
\end{equation*}

Now we partition $\matX_j$ into two parts $\matX_j = [\matX_{1,j}, \matX_{2,j}]$ and $\vecbeta$ compatibly into $\vecbeta = (\vecbeta_1', \vecbeta_2')'$. In this partition, $\matX_{1,j}$ has $p_1$ columns of the design matrix that cannot be written as linear combinations of the columns of $\matZ$. They are typically within-cluster covariates that do not have a correspondent random effect. We can write
\begin{equation*}
\matX_{1,j}=\matZ_j\matW_j+\matE_j	
\end{equation*}
for some $q\times p_1$ matrix $\matW_j$ and some $n_j\times p_1$ matrix $\matE_j$ with $\matE_j'\matZ_j=\matO$. The block of $\matI_{\vecbeta\vecbeta,j}$ for $\vecbeta_1$ can be simplified to
\begin{equation*}
	\matI_{\vecbeta_1\vecbeta_1,j}= \frac{n_j}{\sigma^2}\matS_{EE,n_j}+\matW_j'\left[\frac{\sigma^{2}}{n_j}\matS_{ZZ,n_j}^{-1}+\matPsi\right]^{-1}\matW_j
\end{equation*}
where $\matS_{EE,n_j} = \matE_j'\matE_j/n_j$. For the special situation where the within-cluster covariates are all mean-centered within each cluster and orthogonal to each other, $\matW_j$ is a zero matrix, $\matE =\matX_{1,j}$ and $\matS_{EE,n_j}$ is diagonal.

On the other hand, $\matX_{2,j}$ has the remaining $p_2= p-p_1$ columns of the design matrix that are linear combinations of columns of $\matZ_j$, so we can write $\matX_{2,j}=\matZ_j\matB_j$ for some $q\times p_2$ matrix $\matB_j$. These columns of $\matX_{2,j}$ include
\begin{itemize}
	\item a column of $1$'s for the intercept (which we assume has a random effect),
	\item within-cluster covariates that have both fixed and random effects,
	\item between-cluster covariates (columns of constants within $\matX_{2,j}$), and
	\item interaction effects between between-cluster covariates and within-cluster covariates with random effects.
\end{itemize}
Note that we make the assumption that a between-cluster covariate does not have a random effect, or the individual $\matS_{ZZ,n_j}$ would not be invertible. The correspondent block of $\matI_{\vecbeta\vecbeta,j}$ is reduced to
\begin{align}\label{eq: fixedbeta2}
	\matI_{\vecbeta_2\vecbeta_2,j} = \matB_j'\left[\frac{\sigma^{2}}{n_j}\matS_{ZZ,n_j}^{-1}+\matPsi\right]^{-1}\matB_j
\end{align}
For the special situation where no between-cluster covariate is present and all within-cluster covariates with random effects also have a fixed effect, $\matX_{2,j}$ has the same columns as $\matZ_j$ and $\matB_j$ is an identity matrix.

The off-diagonal block of $\matI_{\vecbeta\vecbeta,j}$ is
\begin{equation*}
	\matI_{\vecbeta_1\vecbeta_2,j}
	=\matW_j'\left[\frac{\sigma^{2}}{n_j}\matS_{ZZ,n_j}^{-1}+\matPsi\right]^{-1}\matB_j
\end{equation*}

If we assume $\frac{1}{N}\sum n_j\matS_{EE,n_j}\rightarrow \matS_{EE}$ (as $J\rightarrow\infty$ and $\min_jn_j\rightarrow\infty$) and $\matS_{EE}$ has full rank, then the block $\matI_{\vecbeta_1\vecbeta_1}=\sum\matI_{\vecbeta_1\vecbeta_1,j}$ has order $O_p(N)$. If $\matPsi$ has full rank, then both $\matI_{\vecbeta_2\vecbeta_2,j}$ and $\matI_{\vecbeta_1\vecbeta_2,j}$ are of constant order, so $\matI_{\vecbeta_2\vecbeta_2}=\sum\matI_{\vecbeta_2\vecbeta_2,j}$ and $\matI_{\vecbeta_2\vecbeta_1}=\sum\matI_{\vecbeta_2\vecbeta_1,j}$ are both of order $O_p(J)$. These suggest that
\begin{equation}\label{eq: fixed}
\log|\matI_{\vecbeta\vecbeta}| = p_2\log(J)+p_1\log(N) +O_p(1)	= p\log(J) +p_1\log(\bar{n}) +O_p(1),
\end{equation}
where $N=J\bar{n}$ ($\bar{n}$ is the average cluster size).

If $S_{EE}$ is singular with rank $\tilde{p}_1<p_1$, then we can linearly reparameterize the fixed effects corresponding to $X_{1,j}$ to separate out $p_1-\tilde{p}_1$ fixed effects whose correspondent (linearly transformed) columns of $\matX_{1,j}$ lies in the columns space of $\matZ_{j}$. These fixed effects can combine into $\matX_{2,j}$. The remaining $\tilde{p}_1$ fixed effects results in a $S_{EE}$ matrix with full rank of $\tilde{p}_1$. Note that a reparameterization may change the determinant of the information matrix, but only up to a multiplicative factor of constant order, so the form of BIC is not affected. In light of this discussion, $p_2$ in Equation~\ref{eq: fixed} can be interpreted as the dimension of the intersection space between the column spaces of $\matX_{j}$ and $\matZ_{j}$: $p_2=\dim\left\{ \mathcal{C}(\matX_{j})\cap\mathcal{C}(\matZ_{j})\right\}$, and $p_1$ as $p-p_2$.

\subsection{Information Matrix of Random Effect Parameters}

We now turn to the block of the information matrix for the random effect parameters. The element corresponding to $\sigma^2$ is given by
\begin{equation*}
	\matI_{\sigma^2\sigma^2,j}
	= \half{1}\tr\left\{\partiald{\matV_j}{\sigma^2}\matV_j^{-1}\partiald{\matV_j}{\sigma^2}\matV_j^{-1}\right\}
	= \half{1}\tr(\matV_j^{-2})
\end{equation*}

Here we note:
\begin{align*}
	\matV_j^{-2}
	=&\frac{1}{\sigma^4}\left\{I_{n_j} - \frac{1}{n_j}\matZ_j\matS_{ZZ,n_j}^{-1}\matZ_j'\right\}\\
	&
	+\frac{1}{n_j^3}\matZ_j\matS_{ZZ,n_j}^{-1}\left[\frac{\sigma^{2}}{n_j}\matS_{ZZ,n_j}^{-1}+\matPsi\right]^{-1}\matS_{ZZ,n_j}^{-1}\left[\frac{\sigma^{2}}{n_j}\matS_{ZZ,n_j}^{-1}+\matPsi\right]^{-1}\matS_{ZZ,n_j}^{-1}\matZ_j'
\end{align*}
and obtain
\begin{align*}
\matI_{\sigma^2\sigma^2,j}
	=\frac{n_j-q}{2\sigma^4}+ \frac{1}{2n_j^2}\tr\left\{\left[\frac{\sigma^{2}}{n_j}\matI_{q}+\matPsi\matS_{ZZ,n_j}\right]^{-2}\right\}
	\end{align*}
Note this term is always of order $O(n_j)$, so $\matI_{\sigma^2\sigma^2}=\sum\matI_{\sigma^2\sigma^2,j}=O(N)$.

The block corresponding to $\vecpsi = \vcth(\matPsi)$ is given by
\begin{align}
	\matI_{\vecpsi\vecpsi,j}
	&= \half{1}\left[\partiald{\vct\matV_j}{\vecpsi'}\right]'\left\{\matV_j^{-1}\otimes\matV_j^{-1}\right\}\left[\partiald{ \vct\matV_j}{\vecpsi'}\right]\\
	&=\half{1}\matD_{q}'\left\{\matZ_j\otimes\matZ_j\right\}'\left\{\matV_j^{-1}\otimes\matV_j^{-1}\right\}\left\{\matZ_j\otimes\matZ_j\right\}\matD_{q}\\
	&=\half{1}\matD_{q}'\left\{\left[\matZ_j'\matV_j^{-1}\matZ_j\right]\otimes\left[\matZ_j'\matV_j^{-1}\matZ_j\right]\right\}\matD_{q}\\
	&=\half{1}\matD_{q}'\left\{\left[\matPsi+\frac{\sigma^2}{n_j}\matS_{ZZ,n_j}^{-1}\right]^{-1}\otimes \left[\matPsi+\frac{\sigma^2}{n_j}\matS_{ZZ,n_j}^{-1}\right]^{-1}\right\}\matD_{q}
\end{align}
where $\matD_q$ is the $q^2\times \half{q(q+1)}$ duplication matrix. If $\matPsi$ has full rank, this matrix has constant order, so $\matI_{\vecpsi\vecpsi} = \sum \matI_{\vecpsi\vecpsi,j} = O_p(J)$.

The off-diagonal block is
\begin{align*}
	\matI_{\vecpsi\sigma^2,j}
	&=\half{1}\left[\partiald{\vct\matV_j}{\vecpsi'}\right]'\left\{\matV_j^{-1}\otimes\matV_j^{-1}\right\}\left[\partiald{ \vct\matV_j}{\sigma^2}\right]\\
	& = \half{1}\matD_{q}'\left\{\matZ_j\otimes\matZ_j\right\}'\left\{\matV_j^{-1}\otimes\matV_j^{-1}\right\}\left[\vct\matI_{n_j}\right]\\
	& = \half{1}\matD_{q}'\vct(\matZ_j'\matV^{-2}\matZ_j)\\
	& = \frac{1}{2n_j}\matD_{q}'\vct\left\{\left[\frac{\sigma^{2}}{n_j}+\matS_{ZZ,n_j}\matPsi\right]^{-2}\matS_{ZZ,n_j}\right\}
\end{align*}
If $\matPsi$ has full rank, this off-diagonal block has order $o_p(1)$, so $\matI_{\vecpsi\sigma^2} = \sum \matI_{\vecpsi\sigma^2,j} = o_p(J)$.

If we denote the vector of variance and covariance parameters by $\vecrho = [\vcth'(\matPsi), \sigma^2]'$, then when $\matPsi$ has full rank,
\begin{equation}\label{eq: random}
	\log|I_{\vecrho\vecrho}| = q^*\log(J)+\log(N) +O(1)= (q^*+1)\log(J) + \log(\bar{n}) +O(1)
\end{equation}
where $q^*= q(q+1)/2$ is the number of non-duplicated elements in $\matPsi$.

\subsection{Summary}

When $\matPsi$ has full rank, Equations ~\ref{eqq4}, \ref{eq: logdetA}, \ref{eq: fixed} and \ref{eq: random} result in the following equation:
\begin{equation}
	\log  f(\yy) = f(\yy|\hat\vart) -\frac{1}{2}[K_1\log(N) + K_2\log(J)]+O(1),
\label{eq: bic11}
\end{equation}
where $K_1=p_1+1$, $K_2=p_2+q^*$, $p_1 = p-p_2$, $p_2=\dim\left\{ \mathcal{C}(\matX_{j})\cap\mathcal{C}(\matZ_{j})\right\}$, and $q^*= q(q+1)/2$. It is common to formulate BIC with deviance as follows:
\begin{equation}
	{\rm BIC}_{E} = -2\log f(\yy|\hat\vart) + K_1\log(N) + K_2\log(J) = -2\log f(\yy|\hat\vart) + K_1\log(\bar{n}) + K\log(J), \label{eq: bic1}
\end{equation}
in which the subscript $E$ stands for ``effective sample size.''
Compared to the conventional BIC, the total penalty term in BIC is decomposed into the two terms, (a) $K_1\log(N)$ and (b) $K_2\log(J)$, which can be also decomposed into (a) $K_1\log(\bar{n})$ and (b) $K\log(J)$ in Equation~\ref{eq: bic1}.

\section{BIC in the Presence of Redundant Random Effects}

The derivations in the section above assume that $\matPsi$ has full rank. When $\matPsi$ is rank deficient, $K_1$ and $K_2$ in Equation~\ref{eq: bic1} may deviate from their prescribed values. Now we investigate such situations.

When $\matPsi$ is singular, some of the random effects are redundant. We first study the situation when some of the random effects have zero variance and the remaining random effects have a variance-covariance matrix of full rank. In this case, we write $\matZ_j = [\matZ_{1,j}, \matZ_{2,j}]$, where the $n_j \times q_1$ block $\matZ_{1,j}$ corresponds to the $q_1$ random effects with a variance-covariance matrix $\matPsi_{11}$ of full rank and the $n_j\times q_2$ block $\matZ_{2,j}$ corresponds to the $q_2$ random effects present in the model but not present in the population. We can write
\begin{equation*}
\matZ_{2,j} = \matZ_{1,j}\overline{\matW}_j+\overline{\matE}_j	
\end{equation*}
for some $q_1\times q_2$ matrix $\overline{\matW}_j$ and some $n_j\times q_2$ matrix $\overline{\matE}$ with $\overline{\matE}_j'\matZ_{1,j} = \matO$. Because we assume $\matZ_{j}$ has full rank, the matrix $\matS_{\overline{\matE}\overline{\matE},n_j} = \overline{\matE}_j'\overline{\matE}_j/n_j$ has full rank.

We also compatibly block $\matPsi$ as
\begin{equation*}
	\matPsi =\left[\begin{array}{cc}
		\matPsi_{11}\ \matPsi_{12}\\
		\matPsi_{21}\ \matPsi_{22}	
	\end{array}\right],
\end{equation*}
where $\matPsi_{22}$, $\matPsi_{21}$ and $\matPsi_{12}$ are zero matrices in the population but are still present in the model. With this formulation, we have
\begin{align*}
\matV_j =& \matZ_{1,j}\matPsi_{11}\matZ_{1,j}'+\sigma^2 I_{n_j}\\
\matV_j^{-1} =& \frac{1}{\sigma^2}\left\{I_{n_j} -\frac{1}{n_j}\matZ_{1,j}\matS_{Z_1Z_1,n_j}^{-1}\matZ_{1,j}'\right\}
		+\frac{1}{n_j^2}\matZ_{1,j}\matS_{Z_1Z_1,n_j}^{-1}
		\left[\frac{\sigma^{2}}{n_j}\matS_{Z_1Z_1,n_j}^{-1}+\matPsi_{11}\right]^{-1}
		\matS_{Z_1Z_1,n_j}^{-1}\matZ_{1,j}'\\
\matV_j^{-2} =& \frac{1}{\sigma^4}\left\{I_{n_j} - \frac{1}{n_j}\matZ_{1,j}\matS_{Z_1Z_1,n_j}^{-1}\matZ_{1,j}'\right\}\\
		&+\frac{1}{n_j^3}\matZ_{1,j}\matS_{Z_1Z_1,n_j}^{-1}
		\left[\frac{\sigma^{2}}{n_j}\matS_{Z_1Z_1,n_j}^{-1}+\matPsi_{11}\right]^{-1}\matS_{Z_1Z_1,n_j}^{-1}
		\left[\frac{\sigma^{2}}{n_j}\matS_{Z_1Z_1,n_j}^{-1}+\matPsi_{11}\right]^{-1}\matS_{Z_1Z_1,n_j}^{-1}\matZ_{1,j}'
\end{align*}

Below we reevaluate the order of the blocks of the information matrix $\matI_j$. For the block $\matI_{\vecbeta\vecbeta,j}$ for fixed effects, we still have
\begin{equation}\label{eq: fixed2}
	\log|\matI_{\vecbeta\vecbeta}| = p_2\log(J)+p_1\log(N) +O_p(1)	= p\log(J) +p_1\log(\bar{n}) +O_p(1),
\end{equation}
where $p_1 = p-p_2$ and $p_2=\dim\left\{ \mathcal{C}(\matX_{j})\cap\mathcal{C}(\matZ_{1,j})\right\}$. Note now $\matZ_{1,j}$ replaces $\matZ_j$ in the definition of $p_2$. This means that some columns of $\matX_{2,j}$ now have to move to $\matX_{1,j}$, resulting in a smaller $p_2$ and a larger $p_1$.

The block $\matI_{\rho\rho}$ of parameters in the random effects is more complicated because blocks $\matPsi_{22}$ and $\matPsi_{21}(=\matPsi_{12}')$ of the variance-covariance matrix $\matPsi$ are zero in the population but are still present in the model. These parameters need to be treated separately. We now define $\vecpsi_{rr}=\vcth\matPsi_{rr}$ ($r = 1, 2$) and $\vecpsi_{12}=\vct\matPsi_{12}$. The blocks of $\matI_j$  are given by

\begin{align*}
\matI_{\vecpsi_{rr}\vecpsi_{ss},j}=&
	\half{1}\matD_{q_r}'\left\{
	\left[\matZ_{r,j}'\matV_j^{-1}\matZ_{s,j}\right]\otimes\left[\matZ_{r,j}'\matV_j^{-1}\matZ_{s,j}\right]
	\right\}\matD_{q_s},\quad\quad (r,s = 1,2)\\
\matI_{\vecpsi_{12}\vecpsi_{ss},j}=&
	\half{1}\left\{
	\left[\matZ_{2,j}'\matV_j^{-1}\matZ_{s,j}\right]\otimes\left[\matZ_{1,j}'\matV_j^{-1}\matZ_{s,j}\right]
	+\left[\matZ_{1,j}'\matV_j^{-1}\matZ_{s,j}\right]\otimes\left[\matZ_{2,j}'\matV_j^{-1}\matZ_{s,j}\right] \right\}\matD_{q_s},\quad\quad (s= 1,2) \\
\matI_{\vecpsi_{12}\vecpsi_{12},j}=&
	\half{1}\left(
	 \left[\matZ_{1,j}'\matV_j^{-1}\matZ_{1,j}\right]\otimes\left[\matZ_{2,j}'\matV_j^{-1}\matZ_{2,j}\right]
	+\left[\matZ_{1,j}'\matV_j^{-1}\matZ_{2,j}\right]\otimes\left[\matZ_{2,j}'\matV_j^{-1}\matZ_{1,j}\right]\right.\\
   &\left.+\left[\matZ_{2,j}'\matV_j^{-1}\matZ_{1,j}\right]\otimes\left[\matZ_{1,j}'\matV_j^{-1}\matZ_{2,j}\right]
	+\left[\matZ_{2,j}'\matV_j^{-1}\matZ_{2,j}\right]\otimes\left[\matZ_{1,j}'\matV_j^{-1}\matZ_{1,j}\right]
	\right)\\
\matI_{\vecpsi_{rr}\sigma^2,j} =&
 	\half{1}\matD_{q_r}'\vct(\matZ_{r,j}'\matV^{-2}\matZ_{r,j}),\quad\quad (r = 1,2)\quad\mbox{and}\\
\matI_{\vecpsi_{12}\sigma^2,j} =&
	 \half{1}\vct(\matZ_{2,j}'\matV^{-2}\matZ_{1,j}+\matZ_{1,j}'\matV^{-2}\matZ_{2,j}) \\
\matI_{\sigma^2\sigma^2,j} =&  \frac{n_j-q_1}{2\sigma^4}+ \frac{1}{2n_j^2}\tr\left\{\left[\frac{\sigma^{2}}{n_j}\matI_{q_1}+\matPsi_{11}\matS_{Z_1Z_1,n_j}\right]^{-2}\right\}
\end{align*}
where
\begin{align*}
\matZ_{1,j}'\matV_j^{-1}\matZ_{1,j} &= \left[\matPsi_{11}+\frac{\sigma^2}{n_j}\matS_{Z_1Z_1,n_j}^{-1}\right]^{-1}\\
\matZ_{2,j}'\matV_j^{-1}\matZ_{2,j} &= \frac{n_j}{\sigma^2}\matS_{\overline{\matE}\overline{\matE},n_j}+\overline{\matW}_j'\left[\frac{\sigma^{2}}{n_j}\matS_{Z_1Z_1,n_j}^{-1}+\matPsi_{11}\right]^{-1}\overline{\matW}_j\\
\matZ_{2,j}'\matV_j^{-1}\matZ_{1,j} &= \overline{\matW}_j'\left[\frac{\sigma^{2}}{n_j}\matS_{Z_1Z_1,n_j}^{-1}+\matPsi_{11}\right]^{-1}\\
\matZ_{1,j}'\matV_j^{-2}\matZ_{1,j} &=
\frac{1}{n_j}\left[\frac{\sigma^{2}}{n_j}\matS_{Z_1Z_1,n_j}^{-1}+\matPsi_{11}\right]^{-1}\matS_{Z_1Z_1,n_j}^{-1}\left[\frac{\sigma^{2}}{n_j}\matS_{Z_1Z_1,n_j}^{-1}+\matPsi_{11}\right]^{-1}\\
\matZ_{2,j}\matV_j^{-2}\matZ_{2,j}& = \frac{n_j}{\sigma^4}\matS_{\overline{\matE}\overline{\matE},n_j}
+\frac{1}{n_j}\overline{\matW}'\left[\frac{\sigma^{2}}{n_j}\matS_{Z_1Z_1,n_j}^{-1}+\matPsi_{11}\right]^{-1}\matS_{Z_1Z_1,n_j}^{-1}\left[\frac{\sigma^{2}}{n_j}\matS_{Z_1Z_1,n_j}^{-1}+\matPsi_{11}\right]^{-1}\overline{\matW}\\
\matZ_{2,j}\matV_j^{-2}\matZ_{1,j}& = \frac{1}{n_j}\overline{\matW}'\left[\frac{\sigma^{2}}{n_j}\matS_{Z_1Z_1,n_j}^{-1}+\matPsi_{11}\right]^{-1}\matS_{Z_1Z_1,n_j}^{-1}\left[\frac{\sigma^{2}}{n_j}\matS_{Z_1Z_1,n_j}^{-1}+\matPsi_{11}\right]^{-1}
\end{align*}
From these results we can see the information matrix of the random effect parameters has the following order:

\begin{equation*}
\left[\begin{array}{cccc}
	\matI_{\psi_{11}\psi_{11},j} & & &\\
	\matI_{\vecpsi_{12}\vecpsi_{11},j} & \matI_{\vecpsi_{12}\vecpsi_{12},j} & &\\
	\matI_{\vecpsi_{22}\vecpsi_{11},j} & \matI_{\vecpsi_{22}\vecpsi_{12},j} & \matI_{\vecpsi_{22}\vecpsi_{22},j} &\\ \matI_{\sigma^2\vecpsi_{11},j} &\matI_{\sigma^2\vecpsi_{12},j} & \matI_{\sigma^2\vecpsi_{22},j} & \matI_{\sigma^2\sigma^2,j}
\end{array}\right]
=\left[\begin{array}{cccc}
	O_p(1) 	& 			& &\\
	O_p(1)	& O_p(n_j) 	& &\\
	O_p(1)  & O_p(n_j)	& O_p(n_j^2) &\\
	o_p(1)	& o_p(1) 	& O_p(n_j) & O_p(n_j)
\end{array}\right]
\end{equation*}

Summing over the blocks $j = 1,2,\cdots, J$, for the case of equal cluster sizes $n_j = n$, as $n, J\rightarrow \infty$, the log-determinant is
\begin{align*}
	\log|\matI_{\rho\rho}| &= (q_1q_2+1)\log(nJ) + q_2^*\log(n^2J) +q_1^*\log(J) +O_p(1)\\
	& = (q^*+1)\log(J)+(2q_2^*+q_1q_2+1)\log(n) + O_p(1)
\end{align*}
where $q^*= q(q+1)/2$ and $q_r^*= q_r(q_r+1)/2$ ($r=1,2$). Note the dramatic increase in the coefficient for $\log(n)$ compared to Equation~\ref{eq: random}.

Combining the two blocks of the information matrices, we have
\begin{equation}
	{\rm BIC}_{E} = -2\log f(\yy|\hat\vart) + K_1\log(n) + K\log(J), \label{eq: BIC1}
\end{equation}
where $K=p+q^*+1$ is the total number of parameters, $K_1=p_1+1+q_1q_2+2q_2^*$, $p_1 = p-p_2$, $p_2=\dim\left\{ \mathcal{C}(\matX_{j})\cap\mathcal{C}(\matZ_{1,j})\right\}$, and $q_2^*= q_2(q_2+1)/2$.

The discussions above assume that some of the random effects in the model have zero variance in the population and all the remaining random effects are not redundant. It may also happen that none of the random effects has zero variance but a certain linear combination of them does, resulting in rank deficiency of $\matPsi$. In this case, a proper linear transformation of the columns of $\matZ_j$ and the random effects $\vecbeta_j$ can be employed to turn this case into the case we just discussed above. Now in Equation~\ref{eq: BIC1} we have $K_1=p_1+1+q_1q_2+2q_2^*$ with $q_1=\mathrm{rank} \matPsi$, $q_2=q-q_1$, $p_2 =\dim\left\{ \mathcal{C}(\matX_{j})\cap\mathcal{C}(\matZ_{1,j}\matU_1)\right\}$, where the columns of the $q\times q_1$ matrix $\matU_1$ are the eigenvectors of $\matPsi$ corresponding to non-zero eigenvalues.

Although we handled the situation of a singular $\matPsi$ above, more difficult situations may arise where $\matPsi$ is \textit{close to} singular. In these situations the large sample behavior of the information matrix would lie between those discussed in Sections 3 and 4. In particular, the coefficient $K_1$ will be greater than that stipulated below Equation~\ref{eq: bic11} but smaller than that stipulated above Equation~\ref{eq: BIC1}.

An additional practical factor that determines the finite sample performance of BIC$_E$ is the relative size of $\matPsi$ and $\sigma^2\matS_{ZZ,n_j}^{-1}/n_j$. As evident from the majority of equations above, BIC$_E$ as a large sample approximation was derived assuming that $\matPsi$ dominates $\sigma^2\matS_{ZZ,n_j}^{-1}/n_j$. This may not be the case if the cluster size $n_j$ is small or if the error variance $\sigma^2$ is large relative to $\matPsi\matS_{ZZ,n_j}$.

\section{\bf Numerical Demonstration}

In this numerical demonstration, we study the finite sample performance of Equation~\ref{eq: bic1}, in particular how the value $K_1$ in finite samples may deviate from its designated value in Section 3.

\subsection{Study Design}

A two-level random-intercept-slope model with balanced cluster sizes ($n_j=n$) was considered as a data-generating model with the following specification:
\begin{equation}
y_{ij} = \beta + \beta_{1}x_{ij} + \beta_{2}x_{j} + b_{0j} + b_{1j}x_{ij} + \epsilon_{ij},\quad(i = 1, 2, \cdots, n;\  j = 1, 2, \cdots, J) \label{gen}
\end{equation}
where $\beta$, $\beta_{1}$, and $\beta_{2}$ are fixed effects, and the random effects $(b_{0j},b_{1j})'\sim MVN({\bf 0},\Psi)$ where the variance of $b_{0j}$ is $\tau_{0}^{2}$, the variance of $b_{1j}$ is $\tau_{1}^{2}$, and the covariance between $b_{0j}$ and $b_{1j}$ is $\tau_{01}$. The random residual $\epsilon_{ij}$ is assumed to follow $N(0, \sigma^{2})$.

In the data-generating model, parameters were set as follows: $\beta=57.98$, $\beta_{1}=1.93$, $\beta_{2}=-14.57$, $\sigma^{2}=42.78$, $\tau_{0}^{2}=40.20$, $\tau_{1}^{2}=21.58$, and $\tau_{01}=-28.95$. These values were chosen from estimates of a two-level random-intercept-slope model reported in Kreft and De Leeuw (2002, pp. 49--50).
The within-cluster covariate ($x_{ij}$) was generated with a normal distribution with mean 0 and standard deviation 2.07.
The between-cluster covariate ($x_j$), the random effects ($b_{0j}$ and $b_{1j}$), and errors of within-cluster units ($\epsilon_{1j}$, $\epsilon_{2j}$, $\cdots$, $\epsilon_{nj}$) were generated from a joint $(n+3)$-variate normal distribution in such as way such that their sample mean vector and the sample covariance matrix (with denominator $J$) match their population true values. This way the MLEs of parameters will match their population true values and the computer program will yield the information matrix $\matA$ evaluated at the same $\hat\vart$, making it possible to study how $\log|\matA|$ changes with $J$ and $n$ free from sampling variability.

Given the estimates reported in Kreft and de Leeuw (1998, pp. 49--50), the correlation between $b_{0}$ and $b_{1}$ is $-0.983$. As discussed earlier, this near perfect correlation results in a larger coefficient $K_1$ of $\log(n)$ than stipulated under Equation~\ref{eq: bic1}. Thus, for Model A, additional correlation true parameter values were considered to show the finite sample performance of BIC$_{E}$ under different magnitudes of correlation between a random intercept and a random slope. The selected magnitudes included $-1.0, -0.8, -0.6, -0.4, -0.2$ and $0$.

Furthermore, as discussed above, the magnitude of $\sigma^{2}$ affects the behavior of BIC$_{E}$.
Specifically, a smaller value of $\sigma^{2}$ makes the first term in $(\sigma^2\matS_{ZZ,n_j}^{-1}/n_j+\matPsi)^{-1}$ smaller compared to the constant order second term and thereby BIC$_{E}$ closer to its desired asymptotic value. Thus, in addition to $\sigma^{2}=42.78$ (which is comparable to $\tau_{0}^{2}=40.20$), the two additional magnitudes, $\sigma^{2}=10$ and $\sigma^{2}=1$, were considered.
To summarize, 18 sets of true parameters for correlation and $\sigma^{2}$ (6 magnitudes of correlation $\times$ 3 magnitudes of $\sigma^{2}$) were chosen.

In addition to a two-level random-intercept-slope model (called Model A), a two-level random-intercept model (called Model B) was also considered as a data-generating model to show the behavior of BIC$_{E}$ when only a random intercept was modeled. True parameters of Model B were set as those of Model A except that $\tau_{1}^{2}$ and $\tau_{01}$ are not present. As in Model A, the two additional magnitudes, $\sigma^{2}=10$ and $\sigma^{2}=1$, were considered in Model B.

For Models A and B, varying multilevel designs with different cluster sizes $n$ and different numbers of clusters $J$ were considered. The values of $n$ and $J$ were selected based on previous research on MLM (e.g., Geldhof, Preacher, \& Zyphur, 2014).
These include $n=10, 20,$ and $40$, and $J=50, 100, 200$, and $400$. In total, there were 12 ($3 \times 4$) multilevel designs having different combinations of $n$ and $J$.

\subsection{Expected Results}

According to our derivations in Sections 3 and 4, BIC$_{E}$ is a linear function of $\log(n)$ and $\log(J)$. To investigate whether the coefficients of $\log(n)$ and $\log(J)$ are consistent with the empirical values obtained in this demonstration, the two terms of Equation~\ref{eq: logdetA}, $\log|\matI_{\vecbeta\vecbeta}|$ and $\log|\matI_{\rho\rho}|$, are both regressed on $\log(n)$ and $\log(J)$.

For both Models A and B, we expect that the coefficient of $\log(J)$ is always $3$ (the total number of fixed effects) for $\log|\matI_{\vecbeta\vecbeta}|$. The coefficient of $\log(J)$ for $\log|\matI_{\rho\rho}|$ should be the total number of parameters in the covariance matrix. This is $4$ for Model A and is $2$ for Model B.

The coefficient of $\log(n)$ is more complicated. The coefficient $K_1$ of $\log(n)$ was expressed below Equations~\ref{eq: bic1} and ~\ref{eq: BIC1} in terms of the rank of $\matPsi$ and the overlapping number of column-space dimensions between the fixed-effect design matrix $\matX_{j}$ and random-effect design matrix $\matZ_j$ (with redundant random effects removed). In both Models A and B, $\matX_{j}$ is given by
\begin{equation*}
	\matX_j = \left(\begin{array}{ccc}
		1 & x_{ij} & x_{j}\\
		1 & x_{ij} & x_{j}\\
	\vdots & \vdots & \vdots\\
		1 & x_{ij} & x_{j}
	\end{array}	\right)
\end{equation*}
where the three columns correspond to the intercept, the within-cluster covariate and the between-cluster covariate, respectively.

For Model B, $\matZ_j=\mathbf{1}_{n}$ is a single column of 1s. The first and third columns of $\matX_j$ are in its column space but the second column is not. This shows that the coefficient of $\log(n)$ in $\log|\matI_{\vecbeta\vecbeta}|$ should be $1$. The coefficient of $\log(n)$ in $\log|\matI_{\vecrho\vecrho}|$ should also be 1, as shown in Equation~\ref{eq: random}. The second row of Table~\ref{sum} summarizes which parameters count towards which coefficient in the two blocks of the information matrix of Model B. The forms of three versions of BIC are also listed. In practice, the coefficients of $\log(n)$ are expected to be greater than 1 due to the presence of the first term in $(\sigma^2\matS_{ZZ,n_j}^{-1}/n_j+\matPsi)^{-1}$, but they will be closer to 1 as $\sigma^2$ decreases.

Model A has two random effects and the asymptotic behavior of the coefficient of $\log(n)$ depends on the rank of the $2\times 2$ matrix $\matPsi$. When the correlation between the two random effects is not $-1$, no random effect is redundant, and all three columns of $\matX_{j}$ are in the column space of $\matZ_{j}$ (which contains columns 1 and 3 of $\matX_j$), so the coefficient of $\log(n)$ should be zero for $\log|\matI_{\vecbeta\vecbeta}|$. The coefficient of $\log(n)$ for $\log|\matI_{\vecrho\vecrho}|$ should be 1, as in Model B. The first row of Table~\ref{sum} provides a summary for Model A. In practice, the coefficients of $\log(n)$ are expected to be greater than these desired values but the deviations are expected to decrease as $\sigma^2$ decreases. In addition, as the random effect correlation decreases from $0$ to $-1$, $\matPsi$ becomes closer to singular and the coefficients of $\log(n)$ are expected to increase and become closer to their values for a singular $\matPsi$ discussed below.

When the random effect correlation in Model A is $-1$, we have $b_{0j}=-b_{1j}$. In this case we can reparametrize the random effects as $(b_{0j}-b_{1j}, b_{0j}+b_{1j})'$ and the new design matrix can be written as $\matZ_{j} = [\matZ_{1,j},\matZ_{2,j}]$ with $\matZ_{1,j}=(1-x_{1j},1-x_{2j},\cdots,1-x_{nj})'$ and $\matZ_{2,j}=(1+x_{1j},1+x_{2j},\cdots,1+x_{nj})'$. Note that $\matZ_{1,j}$ is in the column space of $\matX_j$, so the overlapping number of dimensions is $p_2=1$, so $p_1=p-p_2=2$, which is the coefficient of $\log(n)$ of $\log|\matI_{\vecbeta\vecbeta}|$. For $\log|\matI_{\vecrho\vecrho}|$, because $q_1=q_2=1$ in this case, the coefficient of $\log(n)$ is $2q_2^*+q_1q_2+1=4$.

\subsection{Results}

Figure 1 presents results for Model A.
For a perfect correlation of $-1$, the regression coefficients of $\log(n)$ (circles for $\sigma^{2}=1$, triangles for $\sigma^{2}=10$, and squares for $\sigma^{2}=42.78$ in Figure 1) for fixed and random effects were close to expected values of $K_1=2$ and $K_1=4$, respectively.
As expected, it is observed empirically that the regression coefficients of $\log(n)$ decrease as the random effect correlation and $\sigma^{2}$ decrease. They become close to their desired values for a model with $\matPsi$ of full rank (blue lines in Figure 1).
For both the fixed and random effects in Model A, the regression coefficients of $\log(J)$ coincide with their desired values of $3$ and $4$.

For fixed effects in Model B, the regression coefficients of $\log(n)$ were 1.323 (standard error [SE]=0.121), 1.051 (SE=0.206), and 1.076 (SE=0.126) for $\sigma^{2}=42.78$, $\sigma^{2}=10$, and $\sigma^{2}=1$, respectively.
These regression coefficients were close to the derived value of $K_1=1$.
In addition, the regression coefficients of $\log(J)$ were 2.994 (SE=0.089), 2.928 (SE=0.150), and 2.987 (SE=0.092) for $\sigma^{2}=42.78$, $\sigma^{2}=10$, and $\sigma^{2}=1$, respectively,
which were closed to the derived value of $K_1+K_2=3$ for the fixed effects in Model B.
For random effects in Model B, the regression coefficients of $\log(n)$ were 1.166 (SE=0.010), 1.084 (SE=0.006), and 1.060 (SE=0.004) for $\sigma^{2}=42.78$, $\sigma^{2}=10$, and $\sigma^{2}=1$, respectively, which were close to the derived value of $K_1=1$ for the random effects in Model B.
Furthermore, the regression coefficients of $\log(J)$ were 2.000 (SE=0.008, SE=0.004, and SE=0.003 for $\sigma^{2}=42.78$, $\sigma^{2}=10$, and $\sigma^{2}=1$, respectively) and its expected value ($K_1+K_2=2$) were the same.
To summarize, these results show the consistency between the derived effective sample size values and the empirical values.

\section{\bf Empirical Data Illustration}

In this section, we illustrate the use of BIC$_{E}$ using an empirical data set. The empirical data set, {\tt popular.dat}, was from Chapter 2 of Hox (2010) and it can be freely downloaded from {\tt http://joophox.net/mlbook2/DataExchange.zip}.
The {\tt popular} data set includes 2,000 students from 100 classes ($J=100$). Average class size is 20 ($\bar{n}=20$; ranged from 16 to 26, standard deviation$=2.05$). Each student belongs to one class, indicating that students (level 1) are nested within classes (level 2). The dependent variable ({\tt popular}) is a self-rated popularity scale ranging from 0 to 10. Within-cluster covariates include a binary-coded gender variable and continuous self-rated extraversion scores.
For illustrative purposes, the contrast-coded gender variable ({\tt gender}; $-1$=boy, $1$=girl) was chosen in the model below.
Between-cluster covariate chosen was continuous teacher experience in years (${\tt texp}_{j}$).
ICC based on an unconditional random-intercept model is 0.365 ($=0.702/(0.702+1.222)$), which means that 36.5 \% of the variance
of popularity scores is at the class level. Thus, the unconditional random-intercept model was considered as a baseline model (Model 1 in Table \ref{select}).

We consider a two-level random-intercept-slope model (Equation 2.12 in Hox [2010]) based on the following equation:
\begin{equation}
	{\tt popular}_{ij} = \beta + \beta_{1}{\tt gender}_{ij} + \beta_{2}{\tt texp}_{j} + \beta_{3}{\tt gender}_{ij} \times {\tt texp}_{j}  + b_{0j} + b_{1j}{\tt gender}_{ij} + \epsilon_{ij},
\end{equation}
where $\beta$, $\beta_{1}$, $\beta_{2}$ and $\beta_{3}$ are the fixed effects of intercept, within-cluster covariate ${\tt gender}_{ij}$, between-cluster covariate ${\tt texp}_{j}$ and their cross-level interaction,
$b_{0j}$ and $b_{1j}$ are the random intercept and the random slope of within-cluster covariate ${\tt gender}_{ij}$, and $\epsilon_{ij}$ is the random residual. We assume $\epsilon_{ij}\sim N(0, \sigma^{2})$ and $(b_{0j},b_{1j})'\sim MVN({\bf 0},\Psi)$ where the variances of $b_{0j}$ and $b_{1j}$ are $\tau_{0}^{2}$ and $\tau_{1}^{2}$, and their covariance is $\tau_{01}$.

For the two-level random-intercept-slope model, there are six possible fixed effect models to be compared:
\begin{itemize}
	\item   F1[null]: $\beta_{1}=0$, $\beta_{2}=0$, $\beta_{3}=0$
	\item   F2[{\tt gender}]: $\beta_{1}\neq0$, $\beta_{2}=0$, $\beta_{3}=0$
	\item   F3[{\tt gender},{\tt gender}$\times${\tt texp}]: $\beta_{1}\neq0$, $\beta_{2}=0$, $\beta_{3}\neq0$
	\item   F4[{\tt texp}]: $\beta_{1}=0$, $\beta_{2}\neq0$, $\beta_{3}=0$
	\item   F5[{\tt gender},{\tt texp}]: $\beta_{1}\neq0$, $\beta_{2}\neq0$, $\beta_{3}=0$
	\item   F6[{\tt gender},{\tt texp},{\tt gender}$\times${\tt texp}]: $\beta_{1}\neq0$, $\beta_{2}\neq0$, $\beta_{3}\neq0$
\end{itemize}
In addition, there are two possible random effect models to be compared:
\begin{itemize}
	\item V1[random intercept]: $\sigma^{2}\neq0$, $\tau_{0}^{2}\neq0$, $\tau_{1}^{2}=0$
	\item V2[random intercept\&slope]: $\sigma^{2}\neq0$, $\tau_{0}^{2}\neq0$, $\tau_{1}^{2}\neq0$, $\tau_{01}\neq0$
\end{itemize}

The objective of the current analysis is to simultaneously identify the important covariates that correspond to the level-specific fixed and random effects among 12 candidate models ($6$ fixed effect models $\times$ $2$ random effect models summarized in Table \ref{select}) based on BIC$_{E}$.
In Table \ref{select}, parameters counting towards $K_1$ and $K_2$ are listed for illustrative purposes. The following patterns should be noted. First, the fixed intercept parameter ($\beta$) counts towards $K_2$ because it is associated with a random intercept parameter. Second, the fixed effect of a within-cluster covariate {\tt gender}$_{ij}$ ($\beta_{1}$) and its interaction effect with {\tt texp}$_{j}$ count towards $K_1$ in the random-intercept model because their two columns in the design matrix $\matX_{j}$ are not in the column space of the design matrix $\matZ_{j}$, which has a single column of 1s; however, they count towards $K_2$ when the random slope of {\tt gender}$_{ij}$ is present because their two columns in $\matX_{j}$ are now multiples of the column of $\matZ_{j}$ for \texttt{gender}$_{ij}$. Third, the fixed effect of a between-cluster covariate {\tt texp}$_{j}$ is always counted for $K_2$ as its column in $\matX_j$ is a multiple of 1's.

The BIC$_{E}$ is compared with the other two BIC calculations based on a total sample size $N$ and the number of clusters $J$, denoted by BIC$_{N}$ and BIC$_{J}$, respectively.
Note that the intercept exists in all 12 models, but is known to be nonzero.
The BIC is based on the deviance ($-2 \times$ log-likelihood).
The {\tt lmer} function of the {\tt lme4} library (Bates, M{\"a}chler, Bolker, \& Walker, 2015) was used to fit the 12 models and to obtain the deviance.
the ML method (Goldstein, 1986; Longford, 1987) is considered to be the most appropriate estimation method for information criteria (Verbeke \& Molenberghs, 2000).
The problem with the ML method is that it tends to underestimate variance components. This tendency is not limited to the residual variance and gets worse as the number of fixed effects increase (e.g., Verbeke \& Molenberghs, 2000). In such a case, restricted maximum likelihood (REML; Patterson \& Thompson, 1971) is recommended.
In the current applications, for all models we considered, the deviance from ML and REML was the same and variance estimates from ML and REML differed in the second or third decimal point. Therefore, the deviance from ML was used in calculating BICs.

As shown in Table \ref{select}, the 12 candidate models were ranked differently across BIC$_{E}$, BIC$_{N}$, and BIC$_{J}$ for Models 4-5 and Models 9-12.
Of Models 4-5, BIC$_{E}$ increased from Model 4 (F2+V2) to Model 5 (F3+V1), whereas BIC$_{N}$ and BIC$_{J}$ decreased from Model 4 to Model 5.
Of Models 9-12, all three BICs increased from the random intercept models (V1) to the random intercept \& slope models (V2). However, they suggested different fixed models: BIC$_{E}$ and BIC$_{N}$ suggested Model 9 (F5+V1), whereas BIC$_{J}$ suggested Model 11 (F6+V1).
This example presents different possible results by different kinds of BIC (BIC$_{E}$, BIC$_{N}$, and BIC$_{J}$).

%
%

\section{\bf Summary and Discussion}

BIC is provided in popular statistical software for MLM.
However, no consensus exists yet among in the literature or in popular software on the calculation of BIC for MLM, posing a key problem in the selection of fixed and random effects.
In this paper, we derived the BIC's penalty term for both fixed effect and random effect parameters in a two-level nested model.
The proposed BIC$_E$ can be used to select among models that differ in level-specific fixed and random effects in the MLM applications.

In the derived BIC$_{E}$, the total penalty term in BIC is decomposed into two terms: (a) a term with the log of average sample size per cluster whose multiplier involves the overlapping number of dimensions between the column spaces of the random and fixed effect design matrices and (b) the total number of parameters times the log of the total number of clusters. This result is consistent with Pauler (1998)'s results for selecting among models that differ only in their fixed-effect parameters.

In addition, we studied the situation when a variance-covariance matrix of random effects is singular, giving analytical expression of the coefficient of $\log(n)$ in BIC$_E$ in this situation. We found that in this situation the coefficient is greater than its desired value with a nonsingular random effect variance-covariance matrix. This result is particularly useful when comparing models that differ not only in the number of fixed effects but also in the number of random effects, as redundant random effects likely lead to a singular random effect variance-covariance matrix.


Through a numerical demonstration, BIC$_{E}$ was found to behave consistently with our theoretical predictions. Furthermore, the use of BIC$_{E}$ is illustrated using a textbook example for selecting among candidate models that have different fixed and random effects. As illustrated, BIC$_{E}$ can be easily calculated by using deviance from software and by counting $K_1$, $K_2$, and sample sizes.





In this study, BIC$_{E}$ for the two-level nested design is presented.
It is left for a future study to further explore the form of BIC$_{E}$ in  higher-level nested designs and for other complicated multilevel designs (e.g., a cross-classified design, a longitudinal design with correlated errors).


%
%


%
%
%

\newpage
\centerline{\bf References}

\begin{description}

\begin{sloppypar}

\item Bates, D., M{\"a}chler, M., Bolker, B., \& Walker, S. (2015). Fitting linear mixed-effects models using lme4. {\it Journal of Statistical Software, 67,} 1--48. https://doi.org/10.1016/S0378-3758(02)00336-1

\item Berger, J., Ghosh, J. K., \& Mukhopadhyay, N. (2003). Approximations and consistency of Bayes factors as model dimension grows. {\it Journal of Statistical Planning and Inference, 112,} 241--258. https://doi.org/10.1016/S0378-3758(02)00336-1

%

\item Burnham, K. P., \& Anderson, D. R. (2002). {\it Model selection and multimodel inference: A practical information-theoretic approach} (2nd ed.). New York, NY: Springer. https://doi.org/10.1007/b97636

\item Delattre, M., Lavielle, M., \& Poursat, M. A. (2014). A note on BIC in mixed-effects models. {\it Electronic Journal of Statistics, 8,} 456--475. https://doi.org/10.1214/14-EJS890

\item de Bruijn, N. G. (1970). {\it Asymptotic methods in analysis.} Amsterdam: NorthHolland.

\item de Leeuw, J., \& Meijer, E. (2007). Introduction to multilevel analysis.
    In J. de Leeuw \& E. Meijer (Eds.), {\it Handbook of advanced multilevel analysis} (pp. 1--75). New York, NY: Springer. https://doi.org/10.4324/9780203848852

\item Geldhof, G. J., Preacher, K. J., \& Zyphur, M. J. (2014). Reliability estimation in a multilevel confirmatory factor analysis framework. {\it Psychological Methods, 19}, 72--91. https://doi.org/10.1037/a0032138

\item Goldstein, H. (1986). Multilevel mixed linear model analysis using iterative generalized least squares. {\it Biometrika, 73}, 43--56. https://doi.org/10.1093/biomet/73.1.43

\item Goldstein, H. (2003). {\it Multilevel statistical models} (3rd ed.). New York, NY: Oxford University Press [Distributor]. https://doi.org/10.1002/9780470973394

\item Hamaker, E. L., van Hattum, P., Kuiper, R. M., \& Hoijtink, H. (2011). Model selection based on information criteria in multilevel modeling. In J. J. Hox \& J. K. Roberts (Eds.), {\it Handbook for advanced multilevel analysis} (pp. 231--255). Routledge/Taylor \& Francis Group. https://doi.org/10.4324/9780203848852

\item Hox, J. J. (2010). {\it Multilevel analysis: Techniques and applications} (2nd ed.) Mahwah, New Jersey: IEA. https://doi.org/10.4324/9780203852279

\item Hox, J. J., Moerbeek, M., \& van de Schoot, R. (2018). {\it Multilevel analysis: Techniques and applications} (3rd ed.) New York, NY: Routledge. https://doi.org/10.4324/9781315650982

\item IBM Corp. (2020). {\it IBM SPSS Statistics for Windows, Version 27.0.} Armonk, NY: IBM Corp.

\item Jones, R. H. (2011). Bayesian information criterion for longitudinal and clustered data. {\it Statistics in Medicine, 30,} 3050-3056. https://doi.org/10.1080/01621459.1995.10476572

\item Kass, \& Vaidyanathan, S. K. (1992). Approximate Bayes factors and orthogonal parameters, with application to testing equality of two Binomial proportions. {\it Journal of the Royal Statistical Society. Series B, Methodological, 54(1),} 129--144. https://doi.org/10.1111/j.2517-6161.1992.tb01868.x

\item Kass, R. E., \& Raftery, A. E. (1995). Bayes factors. {\it Journal of the American Statistical Association, 90}, 773--795. https://doi.org/10.1080/01621459.1995.10476572


\item Kreft, I. G., \& de Leeuw, J. (1998). {\it Introducing multilevel modeling.} SAGE Publications, Ltd https://dx.doi.org/10.4135/9781849209366

\item Laird, N. M., \& Ware, J. H. (1982). Random-effects models for longitudinal data. {\it Biometrics, 38,} 963--974. https://doi.org/10.2307/2529876

%

\item Longford, N. T. (1987). A fast scoring algorithm for maximum likelihood estimation in unbalanced mixed models with nested random effects. {\it Biometrika, 74}, 817--827. https://doi.org/10.1093/biomet/74.4.817


\item Lorah, J., \& Womack, A. (2019). Value of sample size for computation of the Bayesian information criterion (BIC) in multilevel modeling. {\it Behavior Research Methods, 51,} 440--450. https://doi.org/10.3758/s13428-018-1188-3

\item McCoach, D. B., \& Black, A. C. (2008). Evaluation of model fit and adequacy. In A. A. O'Connell \& D. B. McCoach (Eds.), {\it Multilevel modeling of educational data} (245-272). Charlotte, NC: Information Age Publishing, Inc.



\item Merkle, E. C., You, D., \& Preacher, K. J. (2016). Testing nonnested structural equation models. {\it Psychological Methods, 21}, 151--163. https://doi.org/10.1037/met0000038

\item Molenberghs, G., \& Verbeke G. (2007). Likelihood ratio, score, and Wald tests in a constrained parameter space. {\it The American Statistician, 61}, 22--27. https://doi.org/10.1198/000313007X171322

\item Muth\'{e}n, L. K., \& Muth\'{e}n, B. O. (1998-2015). {\it Mplus user’s guide. Seventh Edition}. Los Angeles, CA: Muth\'{e}n \& Muth\'{e}n. Retrieved from https://www.statmodel.com/download/usersguide/Mplus

\item Patterson, H. D., \& Thompson, R. (1971). Recovery of inter-block information when block sizes are unequal. {\it Biometrika, 58}, 545--554. https://doi.org/10.2307/2334389

\item Pauler, D. K. (1998). The Schwarz criterion and related methods for normal linear models. {\it Biometrika, 85,} 13-27. https://doi.org/10.1093/biomet/85.1.13


\item Raftery, A. E. (1995). Bayesian model selection in social research. {\it Sociological Methodology, 25}, 111-163. https://doi.org/10.2307/271063
%


\item SAS Institute Inc. (2015). {\it SAS/STAT® 14.1 User's Guide}. Cary, NC: SAS Institute Inc.

\item Schwarz, G. (1978). Estimating the dimension of a model. {\it Annals of Statistics, 6}, 461--464. https://doi.org/10.1214/aos/1176344136


\item Self, S. G., \& Liang, K.-Y. (1987). Asymptotic properties of maximum likelihood estimators and likelihood ratio tests under nonstandard conditions. {\it Journal of the American Statistical Association, 82}, 605--610. https://doi.org/10.1080/01621459.1987.10478472


\item Snijders, T. A. B., \& Bosker, R. J. (1999). {\it Multilevel analysis: An introduction to basic and advanced multilevel modeling.} Thousand Oaks, CA: Sage.

\item Snijders, T. A. B., \& Bosker, R. J. (2012). {\it Multilevel analysis: An introduction to basic and advanced multilevel modeling} (2nd ed.) Thousand Oaks, CA: Sage. https://doi.org/10.1080/10705511.2013.797841

\item Stone, M. (1979). Comments on model selection criteria of Akaike and Schwarz. {\it Journal of the Royal Statistical Society (Series B), 41}, 276--278. https://doi.org/10.1111/j.2517-6161.1979.tb01084.x

\item Stram, D. O., \& Lee, J. W. (1994). Variance components testing in the longitudinal mixed effects model. {\it Biometrics, 50}, 1171-1177.  https://doi.org/10.2307/2533455

\item Stram, D. O., \& Lee, J. W. (1995). Corrections: Variance components testing in the longitudinal mixed effects model. {\it Biometrics 51}, 1196.
https://doi.org/10.2307/2533038

\item Tierney, L., \& Kadane, J. B. (1986). Accurate approximations for posterior moments and marginal densities. {\it Journal of the American Statistical Association, 81}, 82--86. https://doi.org/10.1080/01621459.1986.10478240

\item Verbeke G., \& Molenberghs G. (2000). {\it Linear mixed models for longitudinal data.} New York, NY: Springer. https://doi.org/10.1007/978-1-4419-0300-6

\item Vuong, Q. (1989). Likelihood ratio tests for model selection and non-nested hypotheses. {\it Econometrica, 57}, 307--333. https://doi.org/10.2307/1912557

\item Weakliem, D. L. (1999). A critique of the Bayesian information criterion for model selection. {\it Sociological Methods \& Research, 27}, 359--397. https://doi.org/10.1177/0049124199027003002

\item Whittaker, T. A., \& Furlow, C. F. (2009). The comparison of model selection criteria when selecting among competing hierarchical linear models. {\it Journal of Modern Applied Statistical Methods, 8}, 173--193. https://doi.org/10.22237/jmasm/1241136840

\item Wolfinger, R. (1993). Laplace's approximation for nonlinear mixed models. {\it Biometrika, 80,} 791--795. https://doi.org/10.1093/biomet/80.4.791

%
%

\end{sloppypar}

\end{description}

\clearpage
\newpage

\begin{sidewaystable}
\caption{Numerical Demonstration: Summary of Data-Generating Models and Calculations of $K_1$ and $K_2$ in BIC$_{E}$ with Comparisons to BIC$_{N}$ and BIC$_{J}$}
\label{sum}
\tiny
\begin{center}
\begin{tabular}{lllllrrrrrrr} \\ \hline
Model &Fixed               &Random          &Fixed Effects    &                   &    &Random Effects   &                           &&BIC            &          & \\ \cline{4-5} \cline{7-8} \cline{10-12}
      &              &                &Par. for $K_1$ &Par. for $K_2$   &    &Par. for $K_1$ &Par. for $K_2$           &&BIC$_{E}$         &BIC$_{N}$ &BIC$_{J}$  \\  \cline{4-5} \cline{7-8} \cline{10-12}
Model A&$x_{ij}$,$x_{j}$    &intercept,slope &[0]            &$\beta,\beta_{1},\beta_{2}$[3]& &$\sigma^{2}$[1] &$\tau_{0}^{2},\tau_{1}^{2},\tau_{01}$[3]&&$D+[1\log(N)+6\log(J)]$&$D+7\log(N)$  &$D+7\log(J)$ \\
Model B&$x_{ij}$,$x_{j}$    &intercept       &$\beta_{1}$[1] &$\beta,\beta_{2}$[2]          & &$\sigma^{2}$[1]&$\tau_{0}^{2}$[1]   &&$D+[2\log(N)+3\log(J)]$&$D+5\log(N)$&$D+5\log(J)$ \\ \hline
\end{tabular}
\end{center}
{\it Note.} $D$ indicates deviance ($-2logf(\yy|\hat{\vart})$); Numbers in square brackets indicate $K_1$ or $K_2$; $N=nJ$ is the total sample size
\end{sidewaystable}

\begin{sidewaystable}
\caption{Empirical Study: BIC$_{E}$ with Comparisons to BIC$_{N}$ and BIC$_{J}$}
\label{select}
\tiny
\begin{center}
\begin{tabular}{lllrrrrrrr} \hline
Model  &Fixed                                                       &Random                      &Par. for $K_1$                             &Par. for $K_2$                                                            &Deviance  & &BIC$_{E}$   &BIC$_{N}$  &BIC$_{J}$    \\ \cline{4-6}\cline{8-10}
Model 1 &F1[null]                                                    &V1[random intercept]        &$\sigma^{2}$[1]                              &$\beta,\tau_{0}^{2}$[2]                                                     &6327.5    & &6344.2(12)  &6350.2(12) &6341.2(12)   \\
Model 2 &F1[null]                                                    &V2[random intercept\&slope] &$\sigma^{2}$[1]                              &$\beta,\tau_{0}^{2},\tau_{1}^{2},\tau_{01}$[4]                              &5750.5    & &5776.4(10)  &5788.4(10) &5773.4(10)   \\
Model 3 &F2[{\tt gender}]                                            &V1[random intercept]        &$\beta_{1},\sigma^{2}$[2]                    &$\beta,\tau_{0}^{2}$[2]                                                     &5556.3    & &5580.6(5)   &5586.6(5)  &5574.6(5)    \\
Model 4 &F2[{\tt gender}]                                            &V2[random intercept\&slope] &$\sigma^{2}$[1]                              &$\beta,\beta_{1},\tau_{0}^{2},\tau_{1}^{2},\tau_{01}$[5]                    &5551.5    & &5582.0(6)   &5597.0(7)  &5579.0(7)    \\
Model 5 &F3[{\tt gender},{\tt gender}$\times${\tt texp}]             &V1[random intercept]        &$\beta_{1},\beta_{3},\sigma^{2}$[3]          &$\beta,\tau_{0}^{2}$[2]                                                     &5552.1    & &5584.0(7)   &5590.0(6)  &5575.0(6)    \\
Model 6 &F3[{\tt gender},{\tt gender}$\times${\tt texp}]             &V2[random intercept\&slope] &$\sigma^{2}$[1]                              &$\beta,\beta_{1},\beta_{3},\tau_{0}^{2},\tau_{1}^{2},\tau_{01}$[6]          &5549.5    & &5584.8(8)   &5602.8(8)  &5581.8(8)    \\
Model 7 &F4[{\tt texp}]                                              &V1[random intercept]        &$\sigma^{2}$[1]                              &$\beta,\beta_{2},\tau_{0}^{2}$[3]                                           &6303.0    & &6324.4(11)  &6333.4(11) &6321.4(11)   \\
Model 8 &F4[{\tt texp}]                                              &V2[random intercept\&slope] &$\sigma^{2}$[1]                              &$\beta,\beta_{2},\tau_{0}^{2},\tau_{1}^{2},\tau_{01}$[5]                    &5730.0    & &5760.6(9)   &5775.6(9)  &5757.6(9)    \\
Model 9 &F5[{\tt gender},{\tt texp}]                                 &V1[random intercept]        &$\beta_{1},\sigma^{2}$[2]                    &$\beta,\beta_{2},\tau_{0}^{2}$[3]                                           &5528.5    & &5557.4(1)   &5566.4(1)  &5551.4(2)    \\
Model 10&F5[{\tt gender},{\tt texp}]                                 &V2[random intercept\&slope] &$\sigma^{2}$[1]                              &$\beta,\beta_{1},\beta_{2},\tau_{0}^{2},\tau_{1}^{2},\tau_{01}$[6]          &5524.9    & &5560.0(3)   &5578.0(3)  &5557.0(3)    \\
Model 11&F6[{\tt gender},{\tt texp},{\tt gender}$\times${\tt texp}]  &V1[random intercept]        &$\beta_{1},\beta_{3},\sigma^{2}$[3]          &$\beta,\beta_{2},\tau_{0}^{2}$[3]                                           &5523.4    & &5560.0(2)   &5569.0(2)  &5551.0(1)    \\
Model 12&F6[{\tt gender},{\tt texp},{\tt gender}$\times${\tt texp}]  &V2[random intercept\&slope] &$\sigma^{2}$[1]                              &$\beta,\beta_{1},\beta_{2},\beta_{3},\tau_{0}^{2},\tau_{1}^{2},\tau_{01}$[7]&5520.6    & &5560.4(4)   &5581.4(4)  &5557.4(4)    \\  \hline
\end{tabular}
\end{center}
{\it Note.} Numbers in parentheses indicate rank order of the BIC$_{E}$, BIC$_{N}$, and BIC$_{J}$ from the smallest to the largest; Numbers in square brackets for the columns of Par. (parameters) for $K_1$ and Par. for $K_2$ indicate $K_1$ and $K_2$, respectively.
\end{sidewaystable}

\clearpage
\newpage

\begin{table}
\begin{center}
\begin{tabular}{lll} \\ \hline
                      &Fixed Effects   &Random Effects \\ \hline
Reg. Coeff. for $\log(n)$             &\epsfig{figure=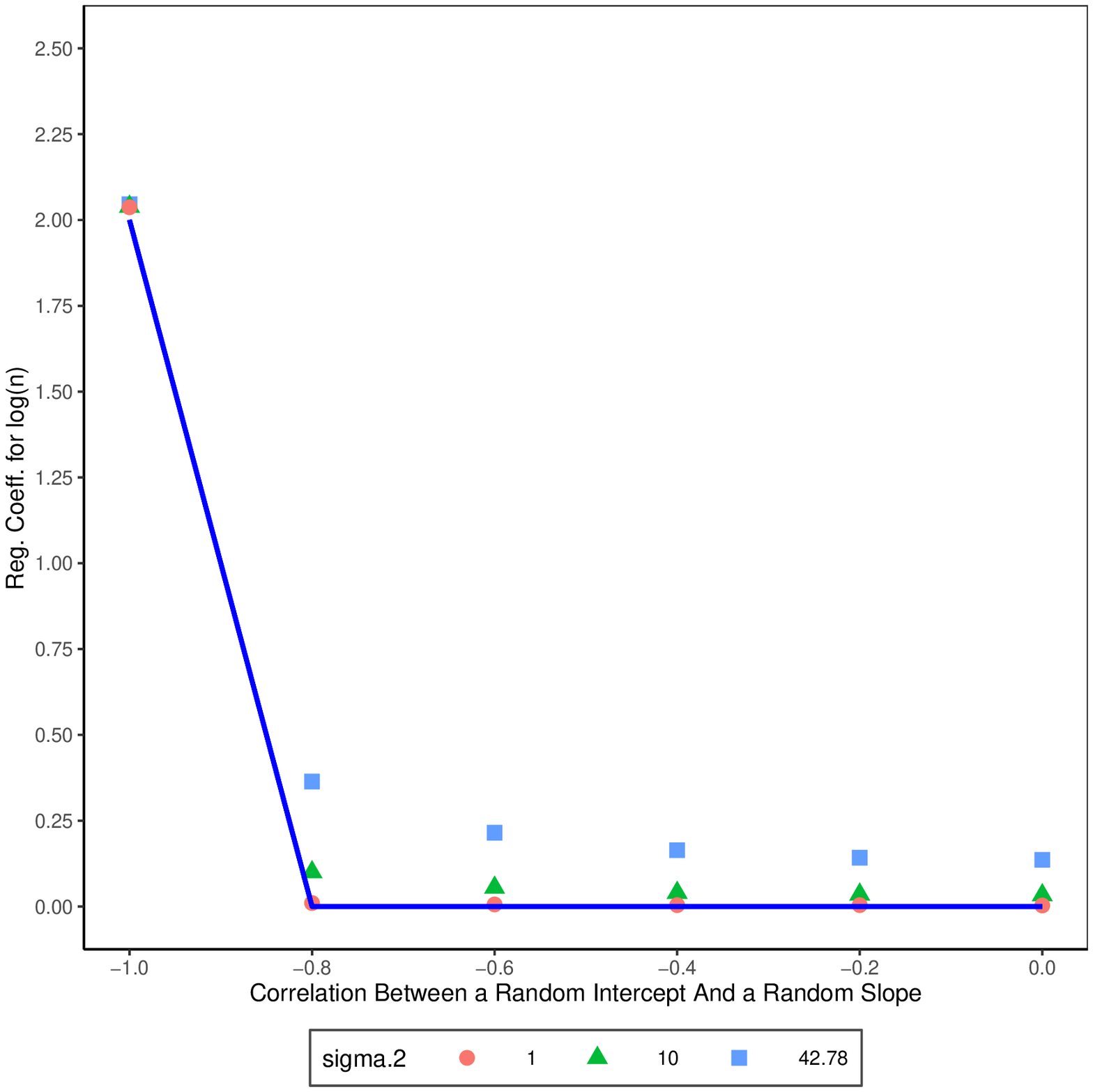,width=0.4\textwidth,totalheight=0.25\textheight}&\epsfig{figure=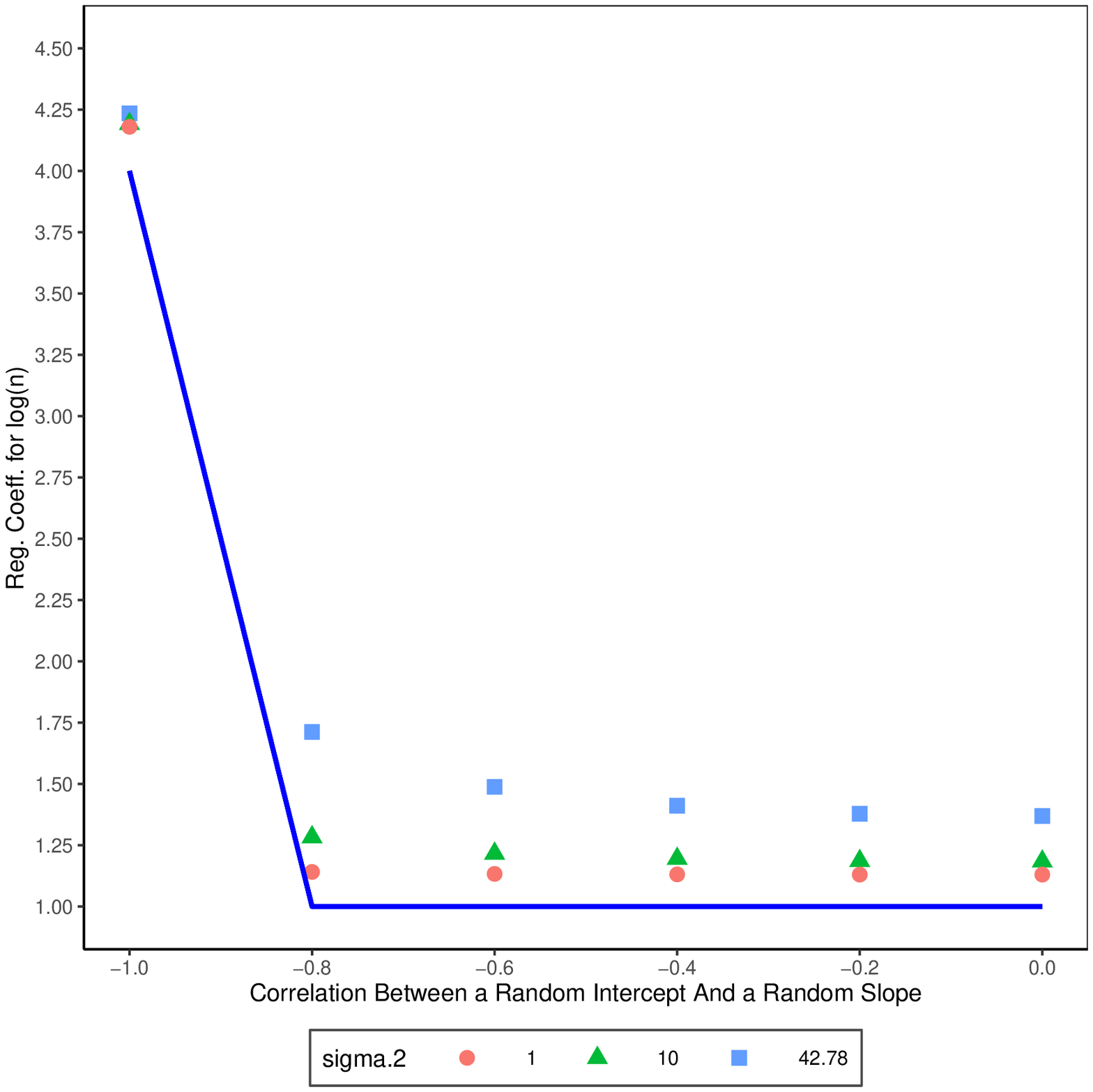,width=0.4\textwidth,totalheight=0.25\textheight} \\ \hline
Reg. Coeff. for $\log(J)$            &\epsfig{figure=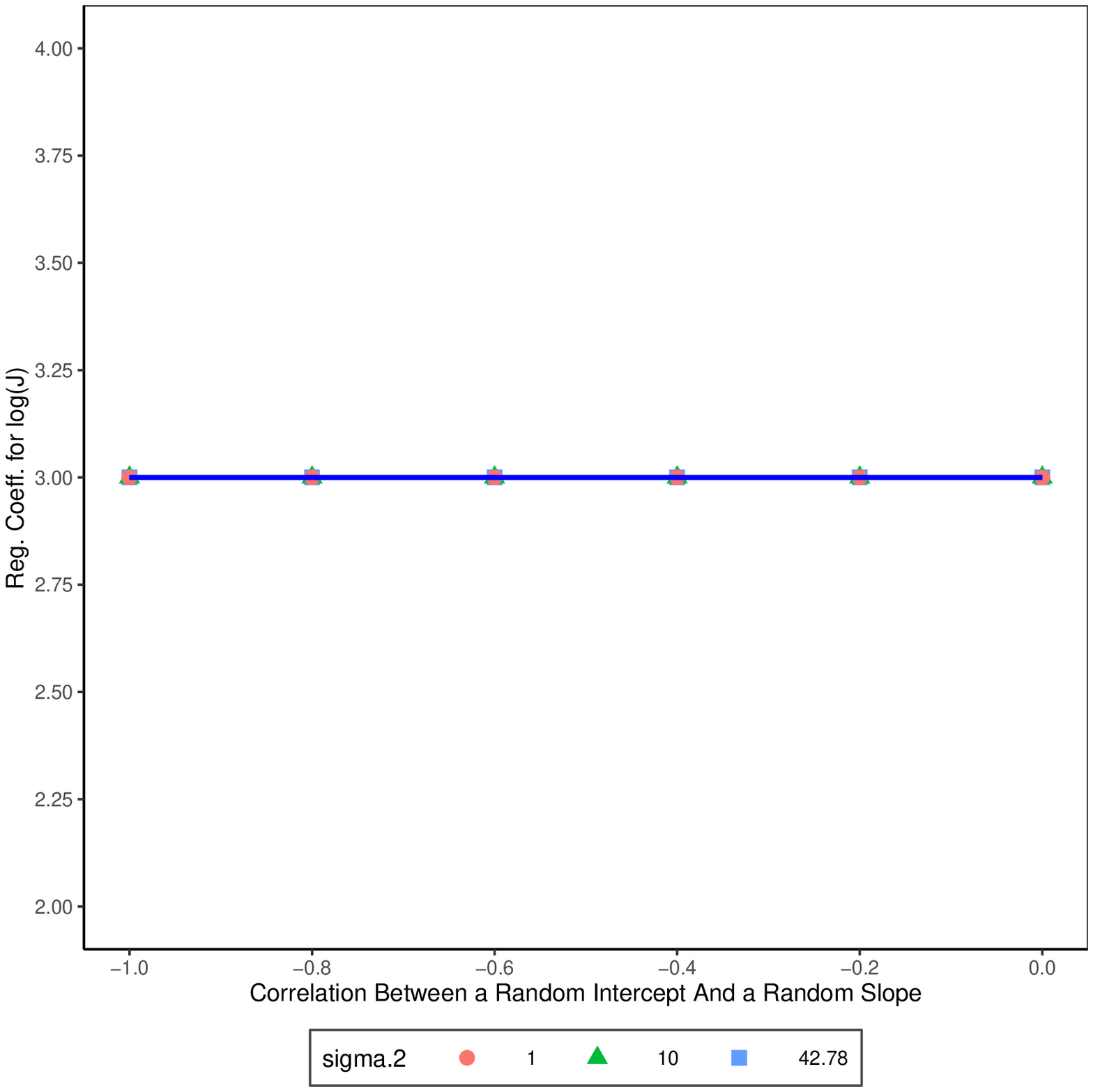,width=0.4\textwidth,totalheight=0.25\textheight}&\epsfig{figure=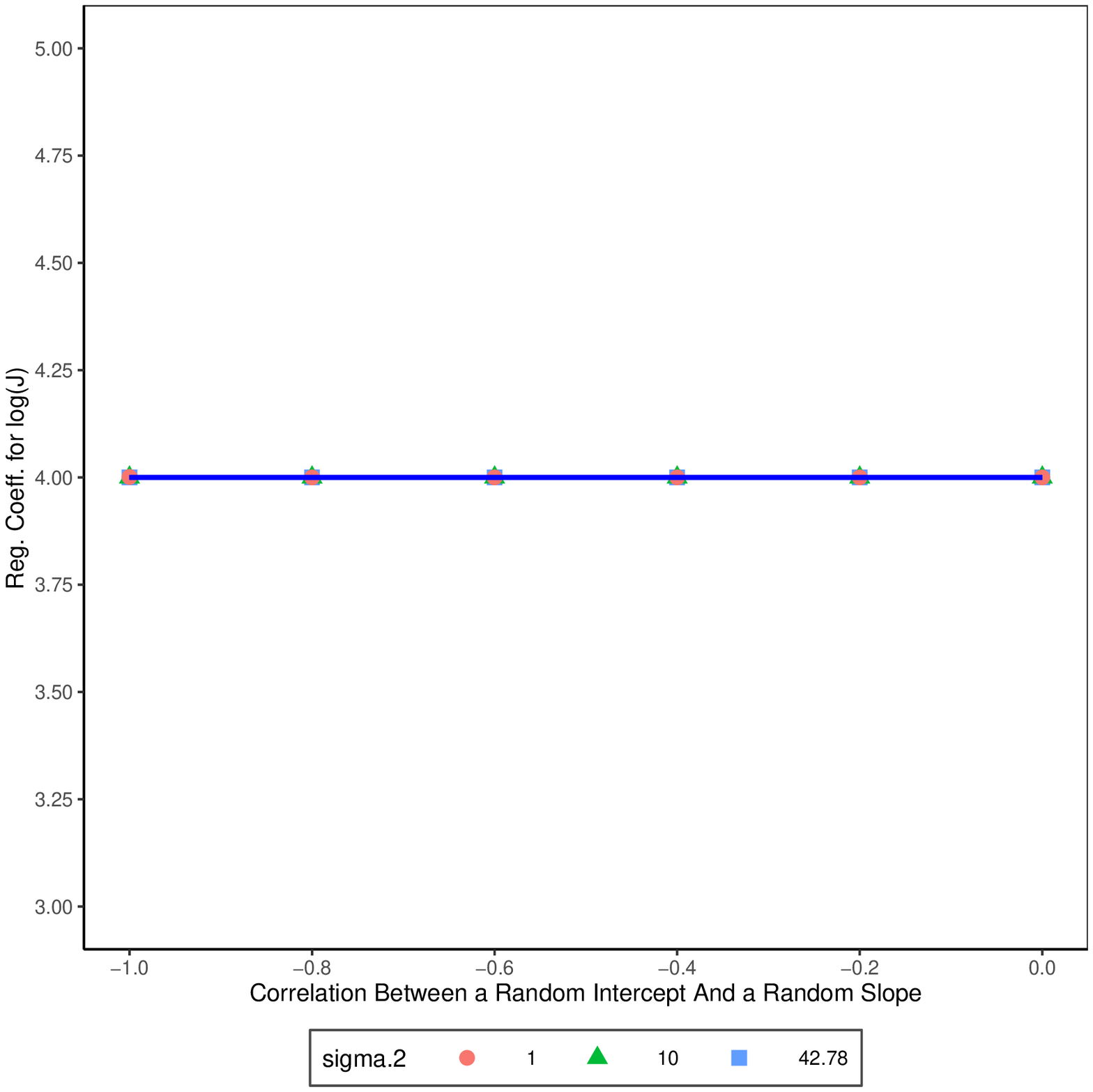,width=0.4\textwidth,totalheight=0.25\textheight}\\ \hline
\end{tabular}
\end{center}
\label{simresult}
{\it Figure 1.} Numerical Demonstration: Results for Model A (Random-Intercept-Slope Model)  \\
{\it Note.} Circles (for $\sigma^{2}=1$), triangles (for $\sigma^{2}=10$), and squares (for $\sigma^{2}=42.78$) indicate empirical values; Lines indicate expected values.
\end{table}

\end{document}